\documentclass[conference]{IEEEtran}

\usepackage{color}
\usepackage[english]{babel}
\usepackage{graphicx}
\usepackage{subcaption}

\begin{document}

\title{Towards dynamic interaction-based reputation models}

\author{\IEEEauthorblockN{Almaz Melnikov, Manuel Mazzara, Victor Rivera, JooYoung Lee}
\IEEEauthorblockA{Innopolis University\\
Russian Federation, Tatarstan republic, Innopolis 420500\\
Emails: a.melnikov@innopolis.ru, m.mazzara@innopolis.ru, \\ v.rivera@innopolis.ru, j.lee@innopolis.ru}
\and
\IEEEauthorblockN{Luca Longo}
\IEEEauthorblockA{School of Computing \\
Dublin Institute of Technology, \\ Dublin, Republic of Ireland\\
Email: luca.longo@dit.ie}
}


\maketitle

% As a general rule, do not put math, special symbols or citations in the abstract
\begin{abstract}
In this paper, we investigate how dynamic properties of reputation can influence the quality of users ranking. Reputation systems should be based on rules that can guarantee a high level of trust and help identifying unreliable units. To understand the effectiveness of dynamic properties in the evaluation of reputation, we propose our own model (DIB-RM) that is based on three factors: forgetting, cumulative, and activity period. In order to evaluate the model, we use data from StackOverflow, which also has its own reputation model. We estimate similarity of ratings between DIB-RM and the StackOverflow model so to check our hypothesis. We use two values to calculate our metric: DIB-RM reputation and $historical$ reputation. We found that $historical$ reputation gives better metric values. Our preliminary results are presented for different sets of values of the aforementioned factors in order to analyze how effectively the model can be used for modeling reputation systems.
\end{abstract}

\IEEEpeerreviewmaketitle

\section{Introduction}
 A feature of interaction-based Internet communities is that direct connections and relationships between users do not have a significant influence on assessing their reputations. Rather, the most influential aspect for such an assessment is the behavior and the activities of the users within a digital community. The computation of user reputation and the assessment of user rating are directly connected because reputation is used for comparing users and at the same time rating is based upon that comparison. Rating systems are used in Internet communities where people communicate with each other, share opinions, information as well as find new contacts. One type of Internet community include web-sites where questioning and answering (Q\&A) is possible. Examples include \emph{Ask.fm} and \emph{Yahoo! answers} that allow users to ask questions on a wide range of topics. Other examples include platforms such as \emph{StackOverflow (SO)} that focuses on more narrow topics as computer science. Q\&A sites are built upon the notion of  community contributions. Here, users generate content by asking specific questions to the community. In turn, other users of the same community can answers them, thus generating peer-reviewed content. The quality of this content depends mainly on the human expertise and knowledge. Hence an open problem is how to assess the level of expertise of users. StackOverflow has its own model for the assessment of the reputation of its users. This is mainly based upon a voting mechanism that allows  users to recommend (like) or disapprove (dislike) the quality of questions or  answers. This mechanism helps to determine the expertise and reputation of each user within the community.  Here, reputation  is an integer value from zero to infinity. As a consequence, users can be ordered and compared by this reputation value.

 This study is focused on the investigation of how dynamic factors - factors that add dynamism to reputation, can be successfully used for rating users. The hypothesis is that dynamic aspects such as past activity, cumulative past knowledge and forgetting (inactivity) can be meaningfully used in computing the reputation of users as well as their trustworthiness in  interaction-based Internet communities. This hypothesis is exploited with the data generated by the StackOverflow platform.\\
 
%Of course, we need to analyze the effectiveness of DIB-RM for reputation evaluation. In this scenario, we compare rating lists that based on StackOverflow and our model reputation values for each user and measure the similarity of them. 
 
 The paper is organized as follows: section 2 describes related works on reputation and trust. Section 3 focuses on the design of a novel model of reputation, called DIB-RM, employing dynamic factors. Section 4 evaluates this model highlighting the impacts of the dynamic factors on the assessment of reputation. Eventually, section 5 concludes the study and presents future work.

\section{Related Work}
\subsection{Reputation and trust}
\begin{figure}[!t]
    \centering
    \includegraphics[width=0.47\textwidth]{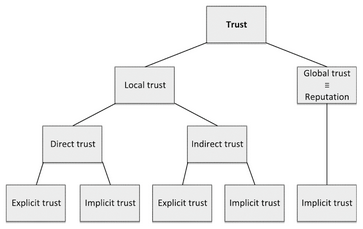}
        \caption{Taxonomy of trust}
        \label{fig:trust}
\end{figure}

Trust can be defined by person's positive or negative expectations of another person's actions. Reputation is a collective measure of trustworthiness based on the referrals or ratings from members of a community.
In \cite{rakoczy2016users}, authors systematize knowledge about trust and reputation. They highlight the problem that many researchers use these terms as equal and therefore they explain and separate them. Authors propose the schema depicted in Fig. \ref{fig:trust} which shows the hierarchy of trust types. Reputation is a type of trust called "Global trust". The first level classification is based on the number of people who participate in trust evaluation:
\begin{enumerate}
\item Local trust - trust which exists between two people.
\item Global trust - trust is the resultant of deposing of the many users' opinions towards one particular user.
\end{enumerate}
Another separation is performed by a method of collecting information:
\begin{enumerate}
\item Explicit - the value is directly given by users.
\item Implicit - the value is based on users activity and interaction, according to available data and made assumptions.
\end{enumerate}
The concept of trust has been investigate thoroughly, and several properties were defined: \textit{context-specific, dynamic, transitive, asymmetric, direction}. As mentioned in \cite{rakoczy2016users}, reputation has only three of them:

\begin{itemize}
\item \textit{Context-specific}. Reputation can be different between the same units of a system in a different scope. Rousseau discussed this specific nature of trust in social and psychological sciences \cite{rousseau1998not}.

\item \textit{Dynamic}. Chang E. \cite{chang2005fuzzy} describes this property in a way that reputation changes on time perspective continuously. Also, new interactions have more influence on reputation value because they are more relevant and important than old ones. A lot of techniques have been invented and they implement this concept \cite{kamvar2003eigentrust}, \cite{zhang2007fine}, \cite{zhou2007powertrust}.

\item \textit{Transitive}. This is the most common property which is widely used in several models. The reputation of a person depends on indirect connections of other people. There are several examples \cite{sabater2002reputation}, \cite{mui2002computational}.
\end{itemize}

Non-commercial trust-based platform have been proposed in the past \cite{MBGDMQN2013}. However, temporal factors have been rarely used as an exclusive factor in the computation of trust.

\subsection{Reputation models}
Nowadays, size of Internet communities increases, more and more people around the world connect to different platforms, such as Facebook, MySpace or Twitter. However, the users meet many problems related to trust. For example, a user needs to know a level of trustworthiness of a service provider or a product supplier before making a choice, or evaluate in new person before accepting his/her request. \cite{hamdi2016computational} Due to the incredible growth of social networks, researchers give their attention to trust and reputation management problems. Measurement of trust in social networks is based on several principles.
Wanita Sherchan separates reputation models into three groups \cite{Sherchan:2013:STS:2501654.2501661}:
\begin{enumerate}
\item Network Structure/Graph-Based models.
\item Interaction-Based models.
\item Hybrid models.
\end{enumerate}
This separation is based on the type of technique which is used in the model.
Models which have network structure use the concept of "Web of trust" or FOAF (Friend-Of-A-Friend). This concept uses "Transitivity" property and direct connections among people to evaluate the trust value between two people. Kutter et Golbeck \cite{kuter2007sunny} invented their model for calculating inference trust in social networks which are called SUNNY. Jiang and Wang \cite{jiang2011swtrust} proposed SWTrust algorithm, it generates a small graph from a big online social network (OSN). Authors in \cite{golbeck2006generating} presented a model which provides a movie recommendation and it is based on an average score of users ratings of films. However, this type of models does not take into account interactions between members. The activity of users and the nature of their communications particularly affect the trust or reputation value.\\

In the previous paragraphs graph based algorithms were mentioned. In contrast to these models, some trust models consider only interactions between system nodes. The name of the group is interaction-based models. Liu et al. actively use in \cite{liu2008predicting} interactions between users in online platforms for predicting trust value. They take into account two groups of parameters: metrics of user's activity with data such as frequency of reviews and ratings and taxonomy of different connections between two users. Kamvar et al. \cite{kamvar2003eigentrust} propose EigenTrust algorithm which performs reputation evaluation on history and state of interactions with the system. It uses aging to differentiate importance of new interactions and old ones.
Hybrid models combine graph structure of system and interactions between units of that system. Anupam et. al provides "SecuredTrust" model \cite{das2012securedtrust} which evaluates trust between multi-agent system units for load balancing and finding malicious agents. This model accounts for a historical information that does not allow malicious units to change their trust value in short period of time. They also implement decreasing of trust value of previous interactions that increases the influence of current activity of the unit. \\

Longo et al. in \cite{longo2007temporal} check hypothesis that temporal based factors, such as activity, frequency, regularity and presence, can be used as an evidence of an entity's trustworthiness. They introduce new algorithm and provide tests on Wikipedia database, there are 12000 users and 94000 articles. They compared prediction metrics with Wikipedia ratings and had satisfactory results. Good prediction rate was 60\%, bad prediction rate was less than 20\%, so this approach can be useful in trust measurement and can be aggregated with more traditional methods. The main drawback of using temporal factors is the amount of information required. A lot of data is needed to evaluate the trustworthiness of article and compare them to each other, because interactions are distributed on time interval where the article exists.
The same author proposed a methodology to continuously align a trust model in force with the changing context within dynamic applications such as forums, blogs, p2p systems. The self-adaptation is reflected in the auto-organisation of the trust function aimed at assessing an agents' trustworthiness \cite{longo2009enabling}.

Adali et al. evaluated trust in a social network, which is based on interaction behavior between two users and propagation of messages of each other \cite{adali2010measuring}. The first feature is called conversation trust, the second - propagative trust. These trust metrics depends only on communication traffic stream, so models are interaction-based ones. Only information about sender, receiver and time parameters of messages were used. Authors investigated the relevance of using this features on Twitter social network database. They divide messages into several sets by proximity of time. These sets are called conversations. Long conversations are also more confidently balanced conversations. Propagative trust is higher if users share messages to third parties.

Several models were designed for trust and reputation evaluation. They solve different problems from implementing recommendation system to reaching the high quality of service and system load balancing. 

\subsection{Research question and hypothesis}
Some researchers improve models by making them more complex and harder in computation to achieve better results. On another hand, some of them try to create more simple models without significant decreasing of results but with better performance. We select the second approach. So, if reputation model based on interactions will give suitable results, implementing a reputation system which needs to store additional data and requires to create and manage new logic is redundant. The research question of this paper is:
\emph{To what extent a model, built upon dynamic interaction factors, can approximate subjective voting of users within the StackOverflow community?
}

%\section{Design and methodology}
\section{Dynamic Interaction Based Reputation Model}
\textit{Trust} can be seen as the amount of interaction among people: the more interaction occurs between two individuals the more one of them trust the other. This makes \textit{trust} very unstable, it actually changes continuously over time \cite{multidisciplinary}. We introduce Dynamic Interaction Based Reputation Model (DIB-RM), model that captures this dynamic property of trust.
%there's a need for a transition from Trust to Reputation

DIB-RM is an interaction-based model among users of a community over time. The model computes a reputation value for each user on the system combining different factors: forgetting factor, the continuous decrease of reputation of an individual; cumulative factor, the importance of users' activities; and activity period factor, the period of time in which the change in the reputation value happened.
%People interact with each other by creating relationships that can be interpreted as trust. The nature of trust is very unstable and changes continuously over time \cite{multidisciplinary}. The dynamic property describes this nature of trust. We introduce Dynamic Interaction Based Reputation Model (DIB-RM) it implements dynamic property of trust.
%DIB-RM is based only on interactions among users over time taking into account the behavior of users and their activities. The model computes reputation for each user on the system combining dynamic factors: forgetting factor, that represents the continuous decrement of reputation; cumulative factor, that regulates the importance of users' activities; and activity period factor that represents the period of time in which the change in the value of reputation happened.

%not sure about the meaning of the next paragraph 
DIB-RM updates the reputation value of each interaction using a fixed number of parameters. This removes the need for storing information about previous interactions. Also, it works in dynamic environments that means a model can update the reputation value of users while they provide some action.
%DIB-RM updates the reputation value of each interaction using a fixed number of parameters. This removes the need for storing information about previous interactions. Also, it works in dynamic environments that means a model can update the reputation value of users while they provide some action.

%The research hypothesis of this study can be formalised as:
%$R.H.:$ Dynamic interaction based model can approximate subjective voting based Stackoverflow reputation model with accuracy more than 85\%.

The following sections explain the assumptions made by the model, the mathematical background for DIB-RM and the metrics used to test the hypothesis. 

%\subsection{Assumptions for DIB-RM}
\subsection{Trust Properties}
DIB-RM is built upon the following two properties of trust behavior:
%Internet community already provides several models for reputation and trust evaluations that can be applied to data. They make different assumptions and are based on different approaches. DIB-RM is built upon the following two properties of trust behavior:
%\begin{enumerate}  
% \item if two person or person and community have not interactions for a long period of time, trust level between them starts to decrease;
% \item if two person interact very frequently and regularly, trust level should increase faster than when they communicate rarely.
% \end{enumerate}

\begin{enumerate}  
\item if two individuals have not interactions for a long period of time, the trust level between them starts to decrease;
\item if two individuals interact very frequently and regularly, the trust level between them should increase faster than when they communicate rarely.
\end{enumerate}

The first property is based on the dynamic property of trust. It requires the continuous change of trust levels over time. The second property comes from \cite{multidisciplinary}. Authors use ``fragile trust" concept to represent that trust levels can change rapidly during short period of time depending on the activity of the user.

\subsection{Model description}
In Internet Communities, interactions occur when there is an activity between two individuals. As an example, in \textit{StackOverflow} there is an interaction between a user and the system when a user posts a question, or between users when a user answers an already posted question. 

Interactions in DIB-RM are modeled by $I_n$
$$I_n=I_{b_n}+I_{c_n}$$
where $n \in 0\ldots N$ is the index of the interaction and $N$ is the total number of interactions of a user. $I_n$ contains a time stamp, when the interaction takes place, and
%what's the meaning of the following sentence?
a value that describes the contribution to the reputation. They can be enumerated by time stamp to form historical chain of user's activity.

%Now we introduce formal description of the model which based on mathematical formulas. We start from definition of interactions' reputation value because it represents the smallest piece of data on that based all calculations.
%Interactions have a time stamp of happening and value that describes a contribution to the reputation. Also, they can be enumerated by increasing creation time to form historical chain of a users' activity.
%$I_n$ - interaction value, $n$ - index of the interaction, $n \in (0, N)$, $N$ - total number of interactions of a user.

Interactions have different effects to the trust value. Each interaction has a basic value $I_{b_n}$.
%Explain the basic value
Depending on the state of communication between a user and the system characterized by activity and frequency, an interaction can be perceived differently. $I_{c_n}$ capture the cumulative part of the interaction, the second property of trust held by DIB-RM. 
%Interactions have different effect to trust level, so they can have different values. Each interaction has a basic value: $I_b$. Depending on state of communication between user and system, which is characterized by activity and frequency, interaction can be perceived differently. We introduce cumulative part of interaction that realizes our second property of trust. Total value of interaction is sum of basic and cumulative parts:
%$$I_n=I_{b_n}+I_{c_n}$$
%$I_{b_n}$ - basic part, $I_{c_n}$ - cumulative part of the interactions' trust value.
It is defined as:
$$I_{c_n}=I_{b_n}*\alpha*(1-\frac{1}{A_{n}+1})$$
where $\alpha$ is the weight of the cumulative part. It shows how big $I_{c_n}$ can grow (if $\alpha = 1$ then $I_{c_n} \in 0\ldots I_{b_n}$). $A_{n}$ is the number of sequential activity periods.

\begin{figure}[!t]
    \centering
    \includegraphics[width=0.3\textwidth]{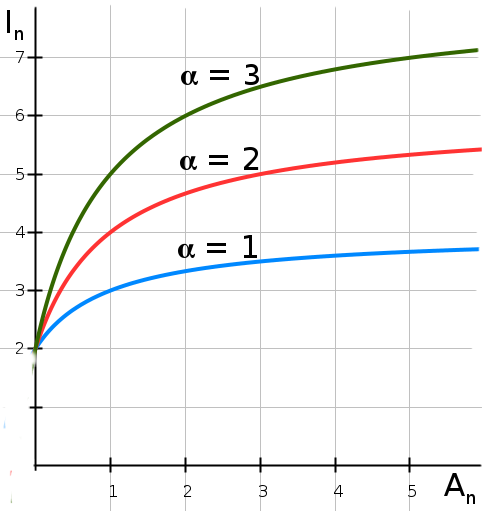}
        \caption{Interaction value graph for different $\alpha$, weight of ``cumulative" effect.}
        \label{fig:interaction}
\end{figure}

Figure \ref{fig:interaction} depicts the dependency of interaction values from a number of activity periods for different weights of $\alpha$ and $I_{b} = 2$. 
%??
So for $\alpha=1$ $I_{c}$ can be maximum 2, for $\alpha=2$ - maximum 4, for $\alpha=3$ - maximum 6.

%mmoved it
%Social communities have different context and features, which affect to properties of a system. One important property is period of time after last user's activity, in which appearance of the next interaction will mean user communicate frequently. We call it activity period and represent as: $t_a$. For example, for Wikipedia it can be one week, when user creates or edits some article, for StackOverflow website it can be one day, when user answers to question or asks.

% moved it 
%Cumulative part of interaction represents, how actively and frequently a user communicate with a system. It is applied for each interaction, those values can be different. That means basic value should be part of cumulative one. We define it as:
%$$I_{c_n}=I_{b_n}*\alpha*(1-\frac{1}{A_{n}+1})$$
%$\alpha$ - weight of the cumulative part, $A_{n}$ - number of activity periods, which come sequentially. 

%moved it
%Weight of the cumulative part, $\alpha$, shows how big $I_{c_n}$ can grow. If $\alpha = 1$, $I_{c_n} \in [0;I_{b_n})$.

%moved it
%The graph depicted in Fig. \ref{fig:interaction} shows dependency of interaction value from number of activity periods for different weights of cumulative part, $\alpha$. Basic value, $I_{b}$, is chosen as 2. So for $\alpha=1$ $I_{c}$ can be maximum 2, for $\alpha=2$ - maximum 4, for $\alpha=3$ - maximum 6.

%moved it
% \begin{figure}[!t]
%     \centering
%     \includegraphics[width=0.3\textwidth]{Interaction_graph}
%         \caption{Interaction value graph for different $\alpha$, weight of "cumulative" effect.}
%         \label{fig:interaction}
% \end{figure}

Social communities have different contexts and features that affect the properties of the system. One of these properties is the frequency of user communication which is defined as the period of time between the last two activities. DIB-RM models this property as $t_a$. As an example, $t_a$ for Wikipedia  can be one week, when a user creates or edits some article whereas for StackOverflow it can be one day, when user answers to a question.
$$\Delta_n=[\frac{t_n - t_{n-1}}{t_a}]$$
is the number of periods between the 2 last interactions. If the difference between $t_n$ and $t_{n-1}$ is less than $t_a$ the number of activity periods will increase by one. It means, the user continues to communicate frequently. 
%Whole part of this difference is got. Otherwise, $A$ will decrease to count of $\Delta_n - 1$.

The final formula for trust is
$$T_n = T_{n-1}*\beta^{\Delta_n} + I_n, \beta \in [0,1]$$
where $\beta$ is the forgetting factor that is chosen by each system individually. If $\beta$ is close to 1, the trust value decreases.
%Trust consists of reduced trust value of previous interaction and the value of current one. The final formula is:
%$$T_n = T_{n-1}*\beta^{\Delta_n} + I_n, \beta \in [0;1]$$
%Forgetting factor of trust value, $\beta$, is a parameter, which chosen for each system individually. If $\beta$ seeks to 1, trust value decrease slower.

\begin{figure}[!t]
    \centering
    \includegraphics[width=0.5\textwidth]{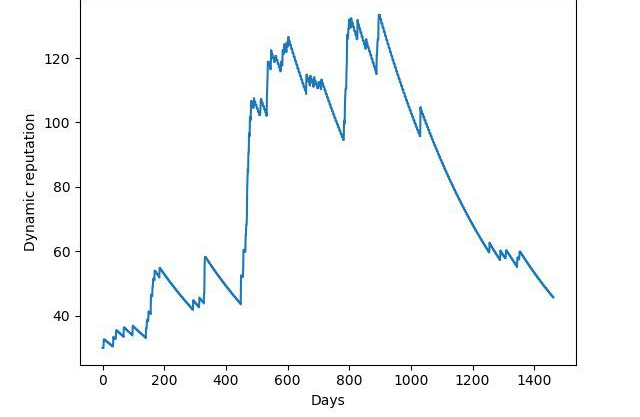}
        \caption{DIB-RM user reputation graph}
        \label{fig:dynamic_reputation}
\end{figure}

%I don't understand the next paragraph
Also, if save DIB-RM reputation values of a user for each day and represent results as a graph, it will look like the line which is depicted in Fig. $\ref{fig:dynamic_reputation}$. Another parameter which can be calculated is the sum of previous reputation values. This parameter is close to a value of an area which is under the graph line. We also use it to compare DIB-RM and StackOverflow model because it accumulates historical information about user's reputation. Even if a user currently has low reputation value but was very active before and done a lot of operations, the sum can be high in comparison with other users. We call this parameter $historical$ reputation.

%Next paragraph is redundant. Everything was said before
%Cumulative factor, $\alpha$, of interaction value and forgetting factor, $\beta$, are implementations of two assumptions that realize dynamic property of trust. Those parameters don't have fixed values, so they can be changed that can lead to different results. 

%Two additional research hypothesis originate from :
%\begin{enumerate}
%\item Increasing of cumulative factor will improve accuracy of approximation 
%\item Decreasing of forgetting factor will improve accuracy of approximation.
%\end{enumerate}
In order to achieve objective results several components should be presented. On one hand it is a reputation model, on another hand, it is data which will be used for evaluations.

\subsection{Metric of approximation}
Reputation values mean nothing in isolation. It is a  relative value used for comparison of users. In general, if the reputation value of user $A$ is higher than the reputation of user $B$, the trustworthiness of user $A$ is also higher. 
%Of course, model designing is just part of research. Another important step is evaluation. To provide strong evidence of efficiency of model and check mentioned hypothesis suitable metric is required. Reputation value does not mean anything, if we consider it apart from others. It is relative value and used for comparison of users. In general if reputation value of user $A$ is higher than reputation of user $B$, trustworthiness of user $A$ also higher. So reputation models directly connected with ranking of users.

%Metric shows similarity between DIB-RM users' rating and StackOverflow rating. Result of comparison of ratings will give information about how DIB-RM approximates StackOverflow voting system and has value of average difference between rating places of users. Formula of metric:

To measure the efficiency of DIB-RM, we applied the model StackOverflow and compared the results to StackOverflow's own rating system. The results of this comparison will give information about how DIB-RM approximates StackOverflow voting system. The metric is defined as

$$\mu_D= 1 - \frac{1}{N^2}*\sum^{N}_{i=1}{(\frac{1}{D}*\sum^{D}_{j=1}{|R_{S_{ij}} - R_{D_{ij}}|})}$$

where $N$ is the number of users, $D$ the number of days between first and last dates, $R_{S_{ij}}$ the StackOverflow reputation value of user $i$ on day $j$ and $R_{D_{ij}}$ is the DIB-RM reputation value of user $i$ on day $j$. $|R_{S_{ij}} - R_{D_{ij}}|$ is the absolute difference between rating places of individual $i$ on particular day $j$. This value shows how close DIB-RM rating is to StackOverflow. Then we calculate average difference of ratings for user $i$ $\frac{1}{D}*\sum^{D}_{j=1}{|R_{S_{ij}} - R_{D_{ij}}|}$ in all-days period. It allows to avoid focusing on one estimation and analyze general behavior of the model. After that average difference of all users is estimated. The last step is subtracting from 1 the average difference, which divided to the number of rating places $N$, gives information about how DIB-RM rating system close to StackOverflow's one.
%First of all, $|R_{S_{ij}} - R_{D_{ij}}|$ absolute difference between rating places of individual $i$ on particular day $j$ is calculated. This value shows how DIB-RM rating place close to StackOverflow rating place. Then we find average difference of ratings for user $i$ $\frac{1}{D}*\sum^{D}_{j=1}{|R_{S_{ij}} - R_{D_{ij}}|}$ in all days period. It allows to avoid focusing on one estimation and analyze general behavior of model. After that average difference of all users is calculated. The last step is subtracting from 1 average difference, which divided to number of rating places $N$, it gives information about how DIB-RM rating system close to StackOverflow's one.

Another approach is measuring rating of users by historical reputation value. The formula of metric remains the same but instead of $R_{D_{ij}}$ (reputation rating place of user $i$ on day $j$) $R_{H_{ij}}$ (historical reputation rating) is used.
$$\mu_H= 1 - \frac{1}{N^2}*\sum^{N}_{i=1}{(\frac{1}{D}*\sum^{D}_{j=1}{|R_{S_{ij}} - R_{H_{ij}}|})}$$

Moreover, error of metric should be estimated to have clear picture of DIB-RM working. If the model has a small error, it gives expected results. Error estimation is performed by calculating standard deviation of metric, $\mu$. For reputation it is $\sigma_D$, for historical reputation it is $\sigma_H$.

\section{Evaluation and Discussion}
We used StackOverflow to evaluate DIB-RM. StackOverflow defines its own users reputation system. We used DIB-RM to evaluate the reputation of users based on interaction. Then, we calculated their difference using the metrics defined in the previous section. The StackOverflow database is available online and can be downloaded from an open access repository. This resource provides $xml$ dumps for all needed files about posts, posts' history, posts' links, comments, users, votes, badges among others. For the computation of DIB-RM and StackOverflow reputations we need posts, comments, users and votes. This is because other entities contain only details about interactions, for example, a post's history stores texts of questions and answers. This dataset includes history of user's activity from September $15^{th}$ 2008, launch day of StackOverflow, to September $14^{th}$ 2012.

%In this research the StackOverflow database is used. By using DIB-RM model we want to know how accurate it can approximate StackOverflow model which based on subjective collective ranking of users. Database of this Internet community is available online and can be downloaded from open access repository. This resource provides $xml$ dumps for all needed files about posts, posts' history, posts' links, comments, users, votes, also badges as well. Computation of DIB-RM and StackOverflow reputations needs only 4 of them: posts, comments, users and votes because other entities contain only details about interactions, for example, post's history store texts of questions and answers. This dataset includes history of user's activity from 15Th September 2008, launch day of StackOverflow, to 14Th September 2012.

\subsection{Analysis of used data}
\begin{figure}[!t]
    \centering
    \includegraphics[width=0.5\textwidth]{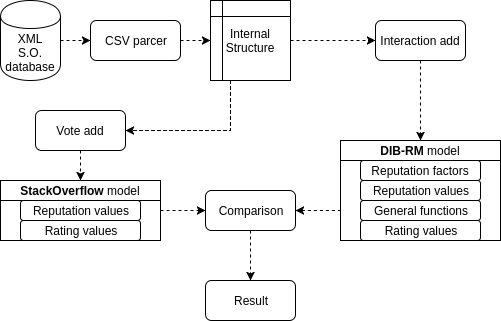}
        \caption{DIB-RM structure.}
        \label{fig:DIB-RM}
\end{figure}

We build a program which internal structure is shown in Fig. \ref{fig:DIB-RM}. Information which is contained in $xml$ files can be represented in a form of a table. Therefore we performed conversion from $ xml$ to $csv$ format because it can be manged by programing tools that we used for creating DIB-RM model. We wrote a parser which optimized to generate output results. It operates only with required fields without converting all file to about$csv$.

The next step is creating internal structure of data from $csv$ files that provides the DIB-RM fast access to information of interactions. By interactions we consider both posts and comments because they show activity of a user and his/her contribution to the system. Post is a general concept of content which users produce. It can be of two types: question or answer. In this paper we do not distinguish types of interactions and assign the same reputation value to them. A typical post tuple is $<PostId, CreationDate, PostTypeId, ParentId, UserId>$, a typical tuple of comment is $<CommentId, CreationDate, UserId>$.

\begin{enumerate}
\item PostId, CommentId - positive integer which represent unique identifier of entity. 
\item CreationDate - date and time when post or comment was created.
\item UserId - positive integer which represents unique identifier of the user who is author.
\end{enumerate}

Those two tuples have similar domains so we can store them together. Sorting of interactions dataset by $(UserId, CreationDate)$ key pair will give historical sequence for each user. We do not add votes as interactions to DIB-RM because the purpose is to compare it with StackOverflow model which is based on voting system.

Vote entities are required to make simulation of StackOverflow model. We created a program which is fully based on rules of calculating users' reputation in StackOverflow. Votes' tuple has the structure $<VoteId, CreatinDate, VoteTypeId, PostId, UserId>$.
\begin{enumerate}
\item VoteId - positive integer which represent unique identifier of vote.
\item PostId - positive integer which represent unique identifier of post. Vote is related to this post.
\item VoteTypeId - positive integer which represent type of vote. It can have a value in the range from 0 to 9.
\item CreationDate - date and time when vote was created.
\end{enumerate}
Each post has a $UserId$ attribute and we can connect a vote with its' recipient and change his reputation.

Total amount of users that we used for computations is 15.000. Between the minimum and maximum StackOverflow's reputations we identified 10 equal intervals and extracted 1.500 users from each group. This method allows to have a representative set of users. During 4 years those users generated 8.630.000 posts, 16.067.000 comments and received 33.269.000 votes.

\subsection{Activity period factor}
The first step of our experimentation is to understand the
importance of the $t_a$ constant (activity period) essential for computing value of "cumulative" part of interaction. We performed a set of computations changing the $t_a$ constant (in days) obtaining ratings difference averages and standard deviation values shown in table \ref{table:1} and in table \ref{table:2}. DIB-RM model has three factors which can be changed: $t_a$, $\alpha$ (cumulative factor), $\beta$ (forgetting factor). Computations are provided with fixed $\alpha=1$ and $\beta=0.99$ for both parameters reputation and historical reputation.

The results which are provided in tables \ref{table:1} and \ref{table:2} show that if $t_a$ (activity period) increases, metric value also increases for both parameters. It comes from the nature of the StackOverflow model which calculates reputation by adding value of a new vote to the sum of previous ones and does not decrease over time. So when $t_a$ increases, reputation value starts to decrease after a longer period of time, users have wider window to interact and increase cumulative part of interactions' reputation value. That means reputation keeps almost the same or increases to the high value of interaction because cumulative part also decreases less often. It makes reputation lose a dynamic property and become static as the StackOverflow model.

Historical reputation does not decrease as StackOverflow reputation also and contains values of all interactions. That is why historical reputation approximates better than the StackOverflow's model reputation value. Values of metric in table \ref{table:2} are between 0.88 and 0.882. That means increasing of $t_a$ does not have significant influence on metric results.

\begin{table}[!t]
\begin{center}
    \begin{tabular}{ | l | l | l | l |}
    \hline
    $\#$ & $t_a$ & $\mu_D$ & $\sigma_D$ \\ \hline
    1 & 1 & 0,8122 & 0,1100  \\  \hline
    2 & 2 & 0,8313 & 0,0936  \\  \hline
    3 & 4 & 0,8510 & 0,0744  \\  \hline
    4 & 8 & 0,8605 & 0,0604  \\  \hline
    \end{tabular}
\end{center}
\caption{Table of reputation metric results for different $t_a$ values}
\label{table:1}
\end{table}

\begin{table}[!t]
\begin{center}
    \begin{tabular}{ | l | l | l | l |}
    \hline
    $\#$ & $t_a$ & $\mu_H$ & $\sigma_H$ \\ \hline
    1 & 1 & 0,8816 & 0,0128  \\  \hline
    2 & 2 & 0,8805 & 0,0026  \\  \hline
    3 & 4 & 0,8813 & 0,0086  \\  \hline
    4 & 8 & 0,8808 & 0,0021  \\  \hline
    \end{tabular}
\end{center}
\caption{Table of historical reputation metric results for different $t_a$ values}
\label{table:2}
\end{table}

\begin{figure}[!t]
\begin{subfigure}{0.24\textwidth}
\includegraphics[width=1\linewidth]{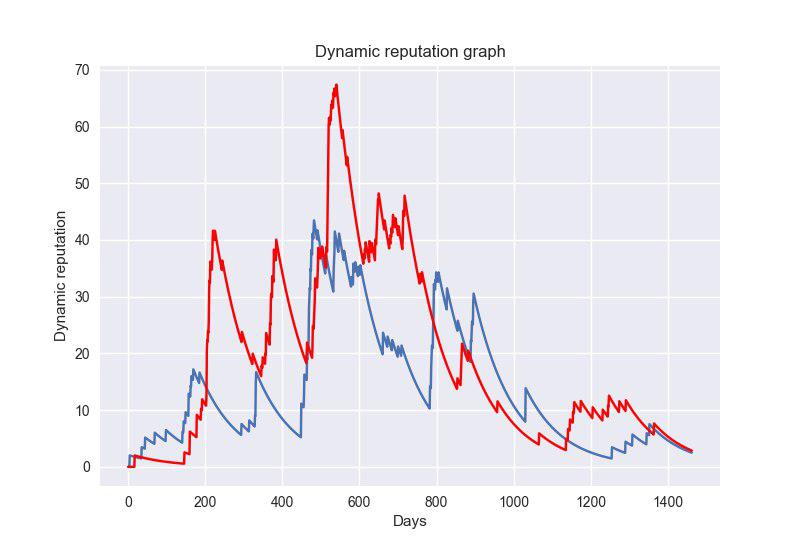} 
\caption{$t_a=1$}
\label{fig:d_rep1}
\end{subfigure}
\begin{subfigure}{0.24\textwidth}
\includegraphics[width=1\linewidth]{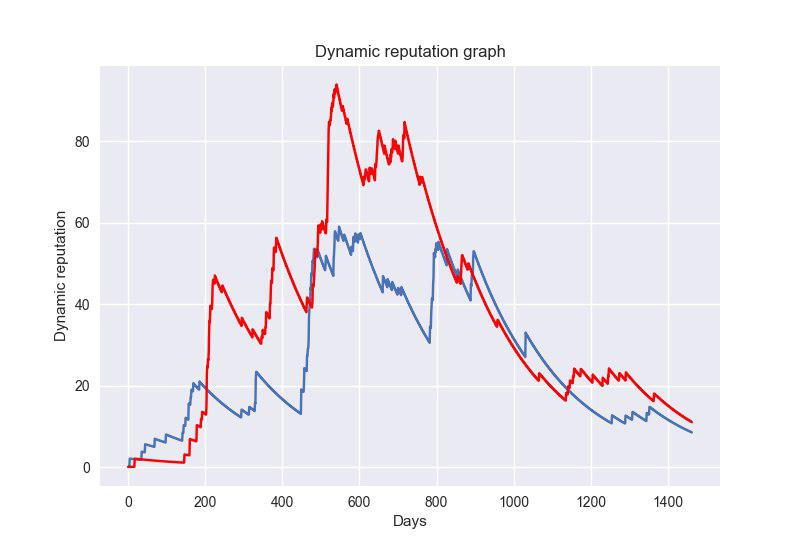} 
\caption{$t_a=2$}
\label{fig:d_rep2}
\end{subfigure}
\begin{subfigure}{0.24\textwidth}
\includegraphics[width=1\linewidth]{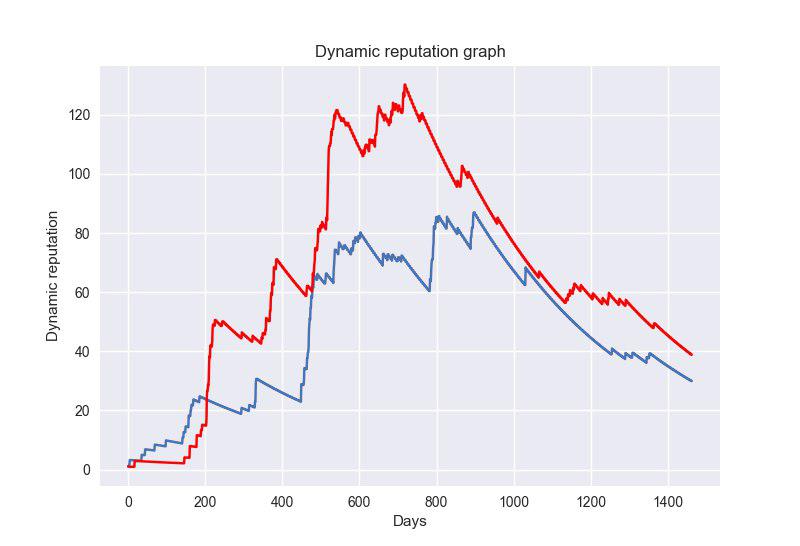} 
\caption{$t_a=4$}
\label{fig:d_rep4}
\end{subfigure}
\begin{subfigure}{0.24\textwidth}
\includegraphics[width=1\linewidth]{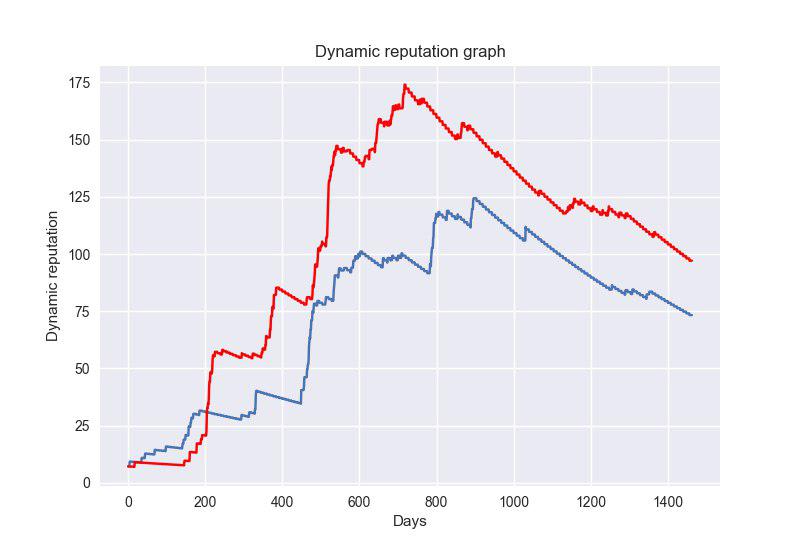} 
\caption{$t_a=8$}
\label{fig:d_rep8}
\end{subfigure}
\caption{Dynamic reputation graph for different $t_a$}
\label{fig:reputation_graph_set}
\end{figure}

Four graphs are depicted in Fig. \ref{fig:reputation_graph_set} for different $t_a$ (activity period) values. They show reputation changes for two users over time. The red line belongs to user with $id = 300$, the blue line to user with $id =235$. Comparison of these lines shows that high value of $t_a$ increases the distance between reputation values. At first sub-graph user $235$ has four times has greater reputation than user $300$. In one period of time in range from 800'th day to 1100'th day blue line is higher than red line. However, on the fourth sub-graph, where $t_a$ parameter equals to 8, the blue line has higher value just at the beginning.

\subsection{Forgetting factor}
In this section we analyze the influence of forgetting factor to metric results and to reputation value. Forgetting factor is used to decrease importance of previous interactions, so new ones have more influence to a reputation. We use two forgetting factor values $\beta=0.99$ and $\beta=0.9$ that means a reputation reduces to 1\% or to 10\% for each activity period. Hence a combination of forgetting factor and activity period factor is also important. The results of computations are presented in tables \ref{table:3} and \ref{table:4}.

We provide metric values for four cases where $\alpha$ is fixed and equals to 1, $t_a$ has two variants, 2 and 8, and $\beta$ equals to 0.99 and 0.90. Increasing the forgetting factor leads to raising of the metric value that means previous interactions' values are also important at reputation evaluation.
If rating of users is built on DIB-RM reputation, changing a $\beta$ value has significant influence to metric. In case of $t_a=2$ $\mu_D$ grows from 0.79 to 0.83 when $t_a=8$ $\mu_D$ grows from 0.81 to 0.86.

\begin{table}[h]
\begin{center}
    \begin{tabular}{ | l | l | l | l | l |}
    \hline
    $\#$ & $t_a$ & $\beta$ & $\mu_D$ & $\sigma_D$ \\ \hline
    1 & 2 & 0.90 & 0,7900 & 0,1285  \\  \hline
    2 & 2 & 0.99 & 0,8313 & 0,0936  \\  \hline
    3 & 8 & 0.90 & 0,8193 & 0,0992  \\  \hline
    4 & 8 & 0.99 & 0,8605 & 0,0604  \\  \hline
    \end{tabular}
\end{center}
\caption{Table of reputation metric results for different $\beta$ values}
\label{table:3}
\end{table}
\begin{table}[h]
\begin{center}
    \begin{tabular}{ | l | l | l | l | l |}
    \hline
    $\#$ & $t_a$ & $\beta$ & $\mu_H$ & $\sigma_H$ \\ \hline
    1 & 2 & 0.90 & 0,8803 & 0,0024  \\  \hline
    2 & 2 & 0.99 & 0,8805 & 0,0026  \\  \hline
    3 & 8 & 0.90 & 0,8808 & 0,0023  \\  \hline
    4 & 8 & 0.99 & 0,8808 & 0,0021  \\  \hline
    \end{tabular}
\end{center}
\caption{Table of historical reputation metric results for different $\beta$ values}
\label{table:4}
\end{table}

\subsection{Cumulative factor}
Cumulative factor $\alpha$ represents the proportion of basic part and cumulative part of interaction. Cumulative part directly depends on the activity of a user. If a user sequentially performs interactions that have an interval between each other less than the activity period, the value of cumulative part increases. We provide evaluation for four cases when $t_a=2$, $\beta=0.99$, $\alpha=\{1, 2, 4, 8\}$. The result values are shown in tables \ref{table:5} and \ref{table:6}.

\begin{table}[!t]
\begin{center}
    \begin{tabular}{ | l | l | l | l | l |}
    \hline
    $\#$ & $\alpha$ & $\mu_D$ & $\sigma_D$ \\ \hline
    1 & 1 & 0,8313 & 0,0936  \\  \hline
    2 & 2 & 0,8441 & 0,0774  \\  \hline
    3 & 4 & 0,8426 & 0,0814 \\  \hline
    4 & 8 & 0,8515 & 0,0723  \\  \hline
    \end{tabular}
\end{center}
\caption{Table of reputation metric results for different $\alpha$ values}
\label{table:5}
\end{table}
\begin{table}[!t]
\begin{center}
    \begin{tabular}{ | l | l | l | l | l |}
    \hline
    $\#$ & $\alpha$ & $\mu_H$ & $\sigma_H$ \\ \hline
    1 & 1 & 0,8805 & 0,0026  \\  \hline
    2 & 2 & 0,8806 & 0,0025  \\  \hline
    3 & 4 & 0,8808 & 0,0030  \\  \hline
    4 & 8 & 0,8808 & 0,0023  \\  \hline
    \end{tabular}
\end{center}
\caption{Table of historical reputation metric results for different $\alpha$ values}
\label{table:6}
\end{table}

\section{Conclusion and Future Work}
In this paper, we investigated the usage of dynamic factors for reputation evaluation. We formally defined reputation model which combines all factors: forgetting factor, cumulative factor and active period factor. Our evaluation was performed in the context of StackOverflow web site. Results based on 4 year history, covering 15.000 users, more than 8.000.000 posts and 33.000.000 votes. We tested our factors and hypothesis by comparing ratings of users that are created by DIB-RM and StackOverflow model. We used two values for creating ratings: reputation and historical reputation. Historical reputation value gave better results, around 88\% similarity between DIB-RM and StackOverflow ratings. Results of evaluation show that this value is resistant to factors' changes, so it allow to adopt model to some environments by selecting different values of factors without decreasing of metric value. We believe that this factors can be used as an evidence of users' trustworthiness in combination with more traditional ones. Our further works will be addressed to determining environments in which context dynamic factors can be used as a strong evidence of trustworthiness.



%\bibliographystyle{IEEEtran}

\bibliographystyle{plain}
\bibliography{bibliography}

% begin{thebibliography}{1}
% end{thebibliography}

\end{document}